\documentclass[floatfix,%
 reprint,
superscriptaddress,
 amsmath,amssymb,
 aps,
pra,
]{revtex4-2}
\usepackage{graphicx}
\usepackage{dcolumn}
\usepackage{bm}
\usepackage{hyperref}
\usepackage{amsfonts}
\usepackage{xcolor}
\usepackage{comment}
\usepackage{booktabs}
\usepackage[normalem]{ulem}
\usepackage{subfigure,subcaption}
\usepackage{makecell}
\usepackage{braket}
\usepackage{float}

\begin{document}

\preprint{APS/123-QED}
\title{Evolution of Entanglement Witness of Dicke State under Noise and Error mitigation}

\author{Tomis Prajapati}
  \email{tomis.1@iitj.ac.in} 
  \affiliation{%
 Department of Physics, Indian Institute of Technology Jodhpur 342030}%
\author{Harsh Mehta}%
 \email{m23iqt011@iitj.ac.in}
 \affiliation{%
 Quantum Information and Computation (IDRP), Indian Institute of Technology Jodhpur 342030}
\author{Shreya Banerjee}
\email{s.banerjee3@exeter.ac.uk}
\affiliation{Department of Physics and Astronomy, University of Exeter, Stocker Road, Exeter-EX4 4QL, United Kingdom}
\affiliation{Center for Quantum Science and Technology, Siksha ’O’ Anusandhan University, Bhubaneswar-751030, Odisha, India}
\author{Prasanta K. Panigrahi}
\email{pprasanta@iiserkol.ac.in}
\affiliation{Center for Quantum Science and Technology, Siksha ’O’ Anusandhan University, Bhubaneswar-751030, Odisha, India}
\affiliation{Department of Physical Sciences, Indian Institute of Science Education and Research Kolkata, Mohanpur-741246, West Bengal, India}

\author{V. Narayanan}
\email{vnara@iitj.ac.in}
 \affiliation{%
 Department of Physics, Indian Institute of Technology Jodhpur 342030}%
\affiliation{%
 Quantum Information and Computation (IDRP), Indian Institute of Technology Jodhpur 342030}
\altaffiliation{}

\begin{abstract}
The experimental verification of multipartite entangled states is essential for advancing quantum information processing. Entanglement witnesses (EWs) provide a widely used and experimentally accessible approach for detecting genuinely multipartite entangled states. In this work, we theoretically derive the entanglement witness for the four-qubit Dicke state and experimentally evaluate it on two distinct IBM 127-qubit Quantum Processing Units (QPUs), namely ibm\_sherbrook and ibm\_brisbane. A negative expectation value of the witness operator serves as a sufficient condition for confirming genuine multipartite entanglement. We report the maximum (negative) values of the witness achieved on these QPUs as $-0.178 \pm 0.009$ and $-0.169 \pm 0.002$, corresponding to two different state preparation protocols. Additionally, we theoretically investigate the effect of various noise channels on the genuine entanglement of a four-qubit Dicke state using the Qiskit Aer simulator. We show the behavior of the EW constructed under the assumption of Markovian and non-Markovian amplitude damping and depolarizing noises, bit-phase flip noise, and readout errors. We also investigate the effect of varying thermal relaxation time on the EW, depicting a bound on the $T_1$ time required for successful generation of a Dicke State on a superconducting QPU.

\textbf{Keywords-} Entanglement Witness, Dicke state, IBM QPUs, Markovian and non-Markovian Noises
\end{abstract}


\maketitle

\section{Introduction}
Quantum entanglement represents a fundamental aspect of quantum mechanics and plays a pivotal role in the field of quantum information science due to the intrinsic correlations it establishes between two or more quantum systems \cite{nielsen_chuang_2010}. This non-classical correlation underpins a wide range of quantum information processing protocols, including quantum teleportation \cite{PhysRevLett.70.1895, PhysRevLett.69.2881}, quantum dense coding \cite{PhysRevA.59.1829,PhysRevA.61.042311}, and quantum cryptography \cite{BENNETT20147,10.1119/1.16243}. While extensive studies have been conducted on bipartite entangled systems, particularly the two-qubit Bell states \cite{PhysRevLett.28.938, PhysRevLett.49.91, Hensen2015,PhysRevLett.115.250401, PhysRevLett.119.010402, PhysRevA.60.R773}, increasing attention is being directed towards multipartite entangled states. Notable examples include the Greenberger-Horne-Zeilinger (GHZ) states \cite{10.1119/1.16243}, W states \cite{PhysRevLett.92.087902}, cluster states \cite{PhysRevA.93.062329}, and Dicke states \cite{PhysRevLett.98.063604, Zhao:15}, each of which exhibits distinct entanglement properties valuable for various quantum technologies. In multipartite systems, distinguishing genuine multipartite entanglement from separable or tri-separable entanglement is essential but challenging. In low-dimensional bipartite systems, such as qubit-qubit or qubit-qutrit setups, the well-known Peres-Horodecki criterion, based on the positivity of the partial transposition (PPT), provides a necessary and sufficient condition for determining separability. However, as the dimensionality increases or the number of subsystems grows, no single, universally applicable criterion for separability exists. Instead, a range of theoretical and experimental techniques have been developed to detect and analyze quantum entanglement. Numerous researchers have contributed to an extensive array of separability conditions and entanglement detection methods, reflecting the complexity and richness of the subject \cite{GUHNE20091, RevModPhys.81.865}. Also, Bell-like inequalities are generally insufficient for this purpose. For \( n > 3 \), even pure \( n \)-partite entangled states can exhibit lower Bell inequality violations than bi-separable states  \cite{PhysRevLett.88.170405}. This shows that we need better methods to identify and study entanglement in these more complex systems.

The most comprehensive and widely adopted method for characterizing multipartite quantum entanglement involves the concept of entanglement witnesses (EWs). EWs, first introduced by Terhal \cite{TERHAL2000319}, can experimentally detect entanglement and completely characterize separable states \cite{PhysRevLett.92.087902, PhysRevLett.114.170504, HORODECKI19961, TERHAL2000319,KOH2015324}. EWs are observables specifically constructed to detect entanglement in quantum systems without requiring full knowledge of the quantum state, which makes them a quicker and easier option compared to measuring the entire state as required in full state tomography. Instead of reconstructing the entire density matrix, EWs rely solely on the expectation value of a specific observable measured in the given quantum state. Geometrically, the set of all quantum states on a Hilbert space of particular dimension forms a convex set. In this context, an EW corresponds to a hyperplane that separates the entangled states from the set of separable states. Mathematically, an EW is a Hermitian operator \( W \) satisfying $Tr(W \rho_{\text{product}})\geq0$, for all pure product states \( \rho_{\text{product}} = |\psi_1\rangle \cdots |\psi_n\rangle \langle \psi_n| \cdots \langle \psi_1| \). A state \( \rho \) is recognized as entangled if and only if $Tr(W \rho) < 0$. While the Peres-Horodecki criterion based on the positivity of the partial transpose (PPT) offers a necessary and sufficient condition for separability in \( 2 \otimes 2 \) and \( 2 \otimes 3 \) systems \cite{HORODECKI19961}, it only acts as a necessary condition of detection of entanglement for general quantum states in higher dimensions \cite{Banerjee2019}. As is well known, the quantum states that are PPT yet entangled are known as bound entangled states, as their entanglement is not distillable. Particularly in these cases, non-decomposable or optimal entanglement witnesses play a crucial role in detecting entanglement.
Further, recent advancements in the classification of multipartite entanglement have demonstrated the utility of EWs in detecting various forms of genuine multipartite entanglement \cite{HORODECKI19961,PhysRevLett.92.087902}. These tools often outperform traditional Bell inequalities, particularly when some prior information about the state is available \cite{Gühne2003}. An EW tailored to identify genuine \( n \) partite entanglement yields non-negative expectation values for all \( (n-1) \) partite entangled states but exhibits negative expectation values for certain genuinely \( n \) partite entangled states.

Entanglement is a very delicate property in quantum information processing and is easily affected by external factors such as noise, decoherence, and unwanted interactions with the environment. These effects can weaken or even completely destroy entanglement, a phenomenon known as entanglement sudden death~\cite{PhysRevA.92.012338, PhysRevA.78.022336, PhysRevA.100.062311}. Thus, it is difficult to observe ideal features like strongly negative entanglement witness (EW) values as well as perfect fidelity in real-world experiments. Thus, it is of genuine interest how EW behaves under realistic decoherence conditions.

In this work, we construct an EW for the four-qubit Dicke state. Further, we investigate the behavior of this EW under the assumption of Markovian depolarizing, amplitude damping, bit-phase flip, readout errors, and thermal relaxation noises. Additionally, we consider non-Markovian quantum channels for amplitude damping and depolarizing noises and present our findings. Finally, we experimentally analyze the EW for the Dicke state generated on two different IBM QPUs \cite{qiskit2024}, with several state-of-the-art error mitigation strategies \cite{PhysRevApplied.20.064027, PhysRevX.11.041039,cai2020,PhysRevA.105.032620}. Expectedly, our results show that the existence of noise tends to reduce the magnitude of EWs in the case of any Markovian noises, often making them less negative or even positive, whereas the non-Markovian channels show oscillations of the constructed EW between negative and positive values.

The remainder of this article is organized as follows: in Sec.~\ref{dickestateEW}, we describe the theoretical construction of the Entanglement Witness for the four-qubit Dicke state. The experimental settings to verify the witness on real quantum processing units are outlined in Sec.~\ref{nonoiseEw}, along with their outcomes with and without error mitigation schemes. The effect of different noise channels on the EW is discussed in Sec.~\ref{noisyEw}. Finally, we conclude the article in Sec.~\ref{conc}.

\vspace{2 mm}

\section{Entaglement Witness operator for four-qubit Dicke state} \label{dickestateEW}
To detect genuine multipartite entanglement in a pure quantum state \(\ket{\psi}\), an EW operator can be defined as:
\begin{equation} \label{eq1}
    \mathcal{W} = \alpha \mathbb{I} - \ket{\psi}\bra{\psi},
\end{equation}

where \( \mathbb{I} \) is the identity operator, and 
\( \alpha = \max_{\ket{\phi} \in B} |\braket{\phi|\psi}|^2 \), 
with \( B \) denoting the set of all bi-separable states. 

This construction ensures that \( \text{Tr}(\mathcal{W} \rho_B) \geq 0 \) for all bi-separable states \( \rho_B \), while a negative expectation value \( \text{Tr}(\mathcal{W} \ket{\psi}\bra{\psi}) < 0 \) confirms that \( \ket{\psi} \) is genuinely multipartite entangled. The coefficient \( \alpha \) reflects the maximal overlap between \( \ket{\psi} \) and any bi-separable state, and is typically evaluated using the Schmidt decomposition \cite{PhysRevLett.92.087902}. For experimental purposes, the witness operator in Eq. \eqref{eq1} is decomposed into local Von-Neumann measurements \cite{PhysRevA.66.062305}.

We consider the four-qubit symmetric Dicke state with two excitations, 
\begin{align}
 \noindent \nonumber \ket{D^{(2)}_4} = \frac{1}{\sqrt{6}} &( \ket{0011} + \ket{0101} + \ket{1001} \\
\noindent +&\ket{0110} + \ket{1010} + \ket{1100} ).
\label{eq2}   
\end{align}

From Eqs.~\eqref{eq1} and \eqref{eq2}, we now construct the corresponding entanglement witness for $\ket{D^{(2)}_4}$.

The local Von-Neumann decomposition of projectors can be expressed in terms of Pauli operators as:
\begin{align*}
 \noindent \ket{0}\bra{0} = \frac{\mathbb{I} + \sigma_z}{2}, \quad \quad &\ket{1}\bra{1} = \frac{\mathbb{I} - \sigma_z}{2},\\
\noindent \ket{0}\bra{1} = \frac{\sigma_x + i\sigma_y}{2}, \quad &\ket{1}\bra{0} = \frac{\sigma_x - i\sigma_y}{2}. 
\end{align*}

This representation facilitates the implementation of the EW through measurable local observables, as, 

\begin{widetext}
\begin{align}
\noindent \nonumber \mathcal{W}_{D^{(2)}_4} &= \frac{2}{3} \mathbb{I} - \left| D^{(2)}_4 \right> \left< D^{(2)}_4 \right|\\
\noindent \nonumber   &= \frac{1}{48} \bigg( 
    29 \mathbb{I}^{\otimes 4} - 3 \big( \sigma_x^{\otimes 4} + \sigma_y^{\otimes 4} + \sigma_z^{\otimes 4} \big) 
    - \big( 
    \sigma_y \sigma_y \sigma_x \sigma_x + \sigma_y \sigma_x \sigma_y \sigma_x + \sigma_x \sigma_y \sigma_y \sigma_x + \sigma_y \sigma_x \sigma_x \sigma_y 
    + \sigma_x \sigma_y \sigma_x \sigma_y + \sigma_x \sigma_x \sigma_y \sigma_y 
    \big) \\
\noindent \nonumber &+ \big( 
    \sigma_z \sigma_z \mathbb{I} \mathbb{I} + \sigma_z \mathbb{I} \sigma_z \mathbb{I} + \mathbb{I} \sigma_z \sigma_z \mathbb{I} + \sigma_z \mathbb{I} \mathbb{I} \sigma_z 
    + \mathbb{I} \sigma_z \mathbb{I} \sigma_z + \mathbb{I} \mathbb{I} \sigma_z \sigma_z 
    \big) 
    + 2 \sum_{i \in \{x, y\}} \bigg( 
    \big( 
    \sigma_z \sigma_z \sigma_i \sigma_i + \sigma_z \sigma_i \sigma_z \sigma_i + \sigma_z \sigma_i \sigma_i \sigma_z \\
\noindent &+ \sigma_i \sigma_z \sigma_i \sigma_z 
    + \sigma_i \sigma_z \sigma_z \sigma_i + \sigma_z \sigma_z \sigma_i \sigma_i 
    \big) 
    - \big( 
    \sigma_i \sigma_i \mathbb{I} \mathbb{I} + \sigma_i \mathbb{I} \sigma_i \mathbb{I} + \mathbb{I} \sigma_i \sigma_i \mathbb{I} + \sigma_i \mathbb{I} \mathbb{I} \sigma_i 
    + \mathbb{I} \sigma_i \mathbb{I} \sigma_i + \mathbb{I} \mathbb{I} \sigma_i \sigma_i 
    \big) \bigg) 
    \bigg).
\label{Ew}   
\end{align}
\end{widetext}

The decomposition of the witness operator present in Eq.(~\ref{Ew})  required 21 measurement settings to characterize the entanglement witness for the four-qubit Dicke state completely. This witness has a positive expectation value for tri-separable, separable, and fully separable states. Here, the theoretical maximum expectation value is given by $Tr[\mathcal{W}_{D^{(2)}_4} \rho_{D^{(2)}_4} ]= -\frac{1}{3}$. Further, using the witness operator, we can directly estimate the fidelity of the four-qubit Dicke state \cite{PhysRevLett.100.200407} as,
\begin{equation}
    F_{|D_{4}^2\rangle}=\frac{2}{3}\mathbb{I}-\mathcal{W}_{D^{(2)}_4}.
\label{fidelity}    
\end{equation}

\section{Experimental verification of  the Entanglement Witness of a four-qubit Dicke state} \label{nonoiseEw}

\subsection{Generation of Dicke State and experimental setting for Witness testing}

To prepare the four-qubit Dicke state, we have taken similar steps as explained in our previous work \cite{tomis}, i.e., we generated four-qubit Dicke states by two methods: a) a unitary gate-based method and b) a state vector-based method. In the unitary gate-based method, a sequence of single-qubit rotation gates is applied along with two-qubit gates to create the desired four-qubit Dicke state provided in Eq. (~\ref{eq2}). The statevector obtained using this method is:

\begin{align}
\nonumber \noindent |D^{(2)}_4\rangle &=
    0.408269\,|0011\rangle + 0.408197\,|0101\rangle \\
\nonumber \noindent &+ 0.408211\,|1001\rangle + 0.408277\,|0110\rangle \\
 \noindent &+ 0.408291\,|1010\rangle + 0.408242\,|1100\rangle .
    \label{statevector_using_gate_based_method}
\end{align} 

The state vector-based method relies on creating a unitary transformation that produces a desired state vector from an input state through isometry. Further, this unitary is decomposed into single-qubit and two-qubit gates. This process works by zeroing out certain amplitudes using rotation and controlled gates. At each step, the statevector evolves by eliminating the contributions arising from a single qubit. Then the inverse of this set of gates is applied to the zero state, preparing the desired statevector \cite{PhysRevA.93.032318}. This method provides a direct approach to achieving the desired state, without the necessity of fine-tuning gate parameters as required in \textbf{gate-based method}, and guarantees that the created quantum state is more precise. The state vector obtained using this method is
\begin{align}
\noindent \nonumber|D^{(2)}_4\rangle=
 \frac{1}{\sqrt{6}} &(|0011\rangle+ |0101\rangle+|0110\rangle \\
\noindent + &|1001\rangle+ |1010\rangle+|1100\rangle).
\label{statevector_using_statevector_based_method}
\end{align}

As can be seen from Eq. (~\ref{Ew}), once the Dicke state is prepared, verification of its entanglement witness requires only 21 measurements \cite{PhysRevLett.98.063604}. However, to estimate the upper limit on the entanglement witness, we performed all 40 measurements. 

For experimental implementation, we utilized two different QPUs freely available through IBM Quantum, i.e., \textit{$ibm\_brisbane$} and \textit{$ibm\_sherbrook$}. For each QPU, we generated the Dicke state using both state preparation methods. To validate the entanglement witness of the Dicke state in each of these four scenarios, we prepared forty circuits, corresponding to forty distinct measurement settings. Every circuit was optimized with optimization level 3 and transpiled into a set of native gates supported by \textit{$ibm\_brisbane$} and \textit{$ibm\_sherbrook$} QPUs. These circuits were then sequentially executed using the \textit{qiskit-ibm-runtime EstimatorV2 Primitive}. Further, we collected our results and constructed the Entanglement Witness. 

Next, we repeated the same procedure in the absence of any error-mitigation strategy, as well as with various mitigation methods, such as dynamical decoupling (DD), twirling (T), and twirled readout error extinction (TREX), as well as combinations of these methods. 

Dynamic decoupling (DD) combats qubit decoherence and noise errors by inserting precisely timed sequences of $\pi$-pulses (e.g., XY4, XX, XpXm) between computational gates. These sequences periodically refocus the qubit's phase, averaging out low-frequency environmental noise, drift, and static biases. Crucially, during vulnerable idle periods in hardware like superconducting qubits (controlled by microwave pulses), DD applies specific pulse sequences that yield a net-zero operation (identity). We implemented the  XX pulse sequence during our executions. While inducing no computational change, these pulses actively counteract decoherence and crosstalk from neighboring controls, thereby extending coherence and reducing errors \cite{PhysRevApplied.20.064027}.

Pauli twirling mitigates coherent gate errors (systematic miscalibrations, unwanted interactions) by conjugating target operations with randomly selected Pauli gates (X, Y, Z) before and after application. This randomization transforms persistent coherent errors, which accumulate predictably and cause biased outcomes, into stochastic Pauli noise channels, effectively symmetrizing the error model towards a depolarizing channel \cite{PhysRevX.11.041039}. By averaging results over many twirled circuit instances, the technique suppresses coherent error bias, yielding expectation values closer to the ideal case, while ensuring no net computational change in circuit \cite{cai2020}.

Twirled Readout Error Extinction (TREX) adapts Pauli twirling’s randomization principle to mitigate measurement errors for Pauli observables. Instead of conjugating gates, it randomly replaces standard measurements with: (1) a Pauli X gate, (2) a measurement, and (3) a classical bit flip. Though computationally identical to ideal measurement, this twirls readout errors to diagonalize the error transfer matrix, mirroring how gate twirling converts coherent errors into stochastic noise. The diagonal structure enables direct error inversion with minimal calibration overhead \cite{PhysRevA.105.032620}.

\subsection{Experimental Results} \label{EWnonoiseresults}

We now present the experimentally measured Entanglement Witness of the four-qubit Dicke state $\ket{D^{(2)}_4}$ prepared on \texttt{ibm\_brisbane} and \texttt{ibm\_sherbrooke}, for each setting of the experiment as described in Sec.~\ref{nonoiseEw}.
Further, we estimate the fidelity of this experimentally prepared state directly from the EW through Eq. (\ref{fidelity}).

\begin{table*}[!hbt]
\centering
\begin{tabular}{|c|c|c|c|c|c|}
\hline
\textbf{Particulars} & \textbf{Ideal simulator} & 
\multicolumn{4}{c|}{\textbf{Gate-based ibm\_sherbrooke}} \\
\cline{3-6}
 & & \textbf{Without Mitigation} & \textbf{DD Mitigation} & \textbf{DD+TREX Mitigation} & \textbf{T+TREX+DD} \\
\hline
\textbf{Witness Operator} & -0.333 & $0.377 \pm 0.081$ & $-0.099 \pm 0.003$ & $-0.141 \pm 0.009$ & $-0.137 \pm 0.077$ \\
\hline
\textbf{Fidelity}         & 0.9999 & $0.289 \pm 0.081$ & $0.766 \pm 0.003$ & $0.808 \pm 0.009$ & $0.804 \pm 0.077$ \\
\hline
\end{tabular}
\caption{Witness operator and fidelity results for a four-qubit Dicke state on \texttt{ibm\_sherbrooke} using the gate-based method.}
\label{gatebased_sherbrook}
\end{table*}

\begin{table*}[!hbt]
\centering
\begin{tabular}{|c|c|c|c|c|c|}
\hline
\textbf{Particulars} & \textbf{Ideal simulator} & 
\multicolumn{4}{c|}{\textbf{Gate-based ibm\_brisbane}} \\
\cline{3-6}
 & & \textbf{Without Mitigation} & \textbf{DD Mitigation} & \textbf{DD+TREX Mitigation} & \textbf{T+TREX+DD} \\
\hline
\textbf{Witness Operator} & -0.333 & $0.290 \pm 0.032$ & $-0.072 \pm 0.035$ & $-0.178 \pm 0.009$ & $-0.140 \pm 0.023$ \\
\hline
\textbf{Fidelity}         & 0.9999 & $0.376 \pm 0.032$ & $0.74 \pm 0.035$   & $0.845 \pm 0.009$ & $0.807 \pm 0.023$ \\
\hline
\end{tabular}
\caption{Witness operator and fidelity results for a four-qubit Dicke state on \texttt{ibm\_brisbane} using the gate-based method.}
\label{gatebased_brisbane}
\end{table*}

\begin{table*}[!hbt]
\centering
\begin{tabular}{|c|c|c|c|c|c|}
\hline
\textbf{Particulars} & \textbf{Ideal simulator} & 
\multicolumn{4}{c|}{\textbf{Statevector-based ibm\_sherbrooke}} \\
\cline{3-6}
 & & \textbf{Without Mitigation} & \textbf{DD Mitigation} & \textbf{DD+TREX Mitigation} & \textbf{T+TREX+DD} \\
\hline
\textbf{Witness Operator} & -0.333 & $-0.007 \pm 0.009$ & $-0.124 \pm 0.014$ & $-0.109 \pm 0.011$ & $-0.169 \pm 0.002$ \\
\hline
\textbf{Fidelity}         & 0.9999 & $0.674 \pm 0.009$ & $0.791 \pm 0.014$ & $0.775 \pm 0.011$ & $0.836 \pm 0.002$ \\
\hline
\end{tabular}
\caption{Witness operator and fidelity results for a four-qubit Dicke state on \texttt{ibm\_sherbrooke} using the statevector-based method.}
\label{statebased_sherbrook}
\end{table*}

\begin{table*}[!hbt]
\centering
\begin{tabular}{|c|c|c|c|c|c|}
\hline
\textbf{Particulars} & \textbf{Ideal simulator} & 
\multicolumn{4}{c|}{\textbf{ibm\_brisbane}} \\
\cline{3-6}
 & & \textbf{Without Mitigation} & \textbf{DD Mitigation} & \textbf{DD+TREX Mitigation} & \textbf{T+TREX+DD} \\
\hline
\textbf{Witness Operator} & -0.333 & $-0.005 \pm 0.036$ & $-0.081 \pm 0.018$ & $-0.166 \pm 0.010$ & $-0.161 \pm 0.014$ \\
\hline
\textbf{Fidelity}         & 0.9999 & $0.672 \pm 0.036$ & $0.748 \pm 0.018$ & $0.832 \pm 0.010$ & $0.827 \pm 0.014$ \\
\hline
\end{tabular}
\caption{Witness operator and fidelity results for a four-qubit Dicke state on \texttt{ibm\_brisbane} using the statevector-based method.}
\label{statebased_brisbane}
\end{table*}

The average values of the EW and the corresponding fidelity for the four-qubit Dicke state prepared using a unitary-gate-based method, obtained using two different QPUs, are summarized in Tables~\ref{gatebased_sherbrook} and~\ref{gatebased_brisbane}. These results show that, even on a \textit{state-of-the-art} quantum processing unit with $127$ qubits, applying error mitigation techniques is essential even for mere observance of genuine multipartite entanglement within only $4$ of the qubits. 

\begin{figure}[!hbt]
    \centering
    \includegraphics[width=\linewidth]{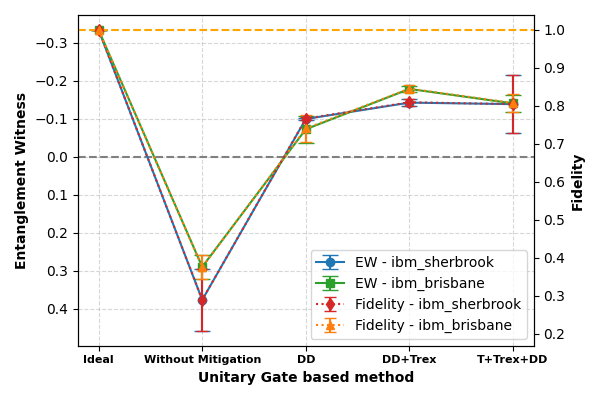}
    \caption{A comparison of the EW and calculated fidelity results for the four-qubit Dicke state prepared by unitary gate-based method on two different QPUs.}
    \label{witness_gatebased_fig}
\end{figure}

Without any error mitigation, no negative EW values were observed on either QPU. However, as we applied more advanced mitigation methods, the results significantly improved on both devices. This improvement is mainly due to the ability of these techniques to reduce gate and hardware-related errors. We found that, for both QPUs, the best results were obtained when both DD and TREX were applied. Under this setting, we achieved the highest (negative) EW values of $-0.141 \pm 0.009$ on \texttt{ibm\_sherbrook} and $-0.178 \pm 0.009$ on \texttt{ibm\_brisbane}. Under the same conditions, we also observed the highest fidelity of $0.808 \pm 0.009$ on \texttt{ibm\_sherbrook} and $0.845 \pm 0.009$ on \texttt{ibm\_brisbane}. A comparison of the EW values and fidelity for the four-qubit Dicke state prepared using the unitary gate-based method is illustrated in Fig.~\ref{witness_gatebased_fig}.

\begin{figure}[!hbt]
    \centering
    \includegraphics[width=\linewidth]{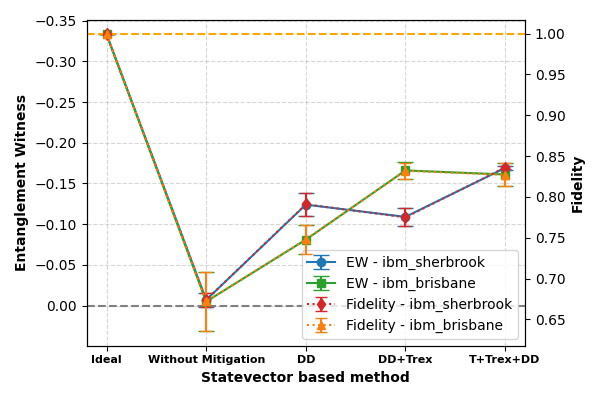}
    \caption{A comparison of the EW and calculated fidelity results for the four-qubit Dicke state prepared by statevector-based method on two different QPUs.}
    \label{fig_witness_statevector}
\end{figure}

For the four-qubit Dicke state prepared using the statevector-based method, the average experimental values of the EW and the corresponding fidelity are presented in Tables~\ref{statebased_sherbrook} and~\ref{statebased_brisbane}. Unlike the gate-based approach, in this case, the entanglement witness attains, albeit very small, but negative values, even without the error mitigation schemes, as can be seen from the Tables~\ref{statebased_sherbrook} and~\ref{statebased_brisbane}. This implies, the Dicke state prepared through this method shows tiny signatures of genuine four-party entanglement, even when no error mitigation technique is applied. This suggests that the state prepared using the statevector-based method is inherently more stable and less affected by noise. This is further attested by the higher fidelity of the Dicke state prepared using this method.

When all three error mitigation techniques, Pauli twirling (T), TREX, and DD, were applied on \texttt{ibm\_sherbrook}, we obtained the most negative EW value of $-0.169 \pm 0.002$ and the highest fidelity of $0.836 \pm 0.002$. On the \texttt{ibm\_brisbane}, the best results were achieved with only DD and TREX applied, yielding an EW of $-0.166 \pm 0.010$ and a fidelity of $0.832 \pm 0.010$. We note that with the application of the error mitigation schemes, the Dicke state, prepared with both state preparation methods, attains similar values of EW across QPUs.
A comparison of the EW values and fidelity for the Dicke state prepared using this method is shown in Fig.~\ref {fig_witness_statevector}, highlighting the overall improved performance of the state-vector-based approach.

\section{Entanglement Witness of Dicke State under assumption of Noise}\label{noisyEw}
As is well known, EWs are a useful tool for detecting entanglement in multipartite quantum systems. However, their effectiveness can be significantly affected by different types of noise that often occur in real quantum hardware. In this section, we explore how the EW of $\ket{D^{(2)}_4}$ behaves under various noise channels. We consider both Markovian and non-Markovian versions of amplitude damping and depolarizing noise, as well as Markovian bit-phase flip noise. We also consider the effect of readout errors and Thermal Relaxation time on the entanglement witness. 

\subsection{Amplitude Damping Noise}

\subsubsection{Markovian Amplitude Damping}
Markovian amplitude damping (AD noise) is a fundamental noise process in quantum computing caused by the irreversible loss of energy from a qubit to its environment \cite{Preskill2018quantumcomputingin}. This phenomenon primarily drives transitions from the excited state (\(\ket{1}\)) to the ground state (\(\ket{0}\)), mimicking spontaneous emission in atomic systems. AD noise is asymmetric in nature, i.e.,the \(\ket{0}\) state remains unaffected, while the \(\ket{1}\) state decays probabilistically.

\begin{figure}[h!]
    \centering
    \includegraphics[width=1\linewidth]{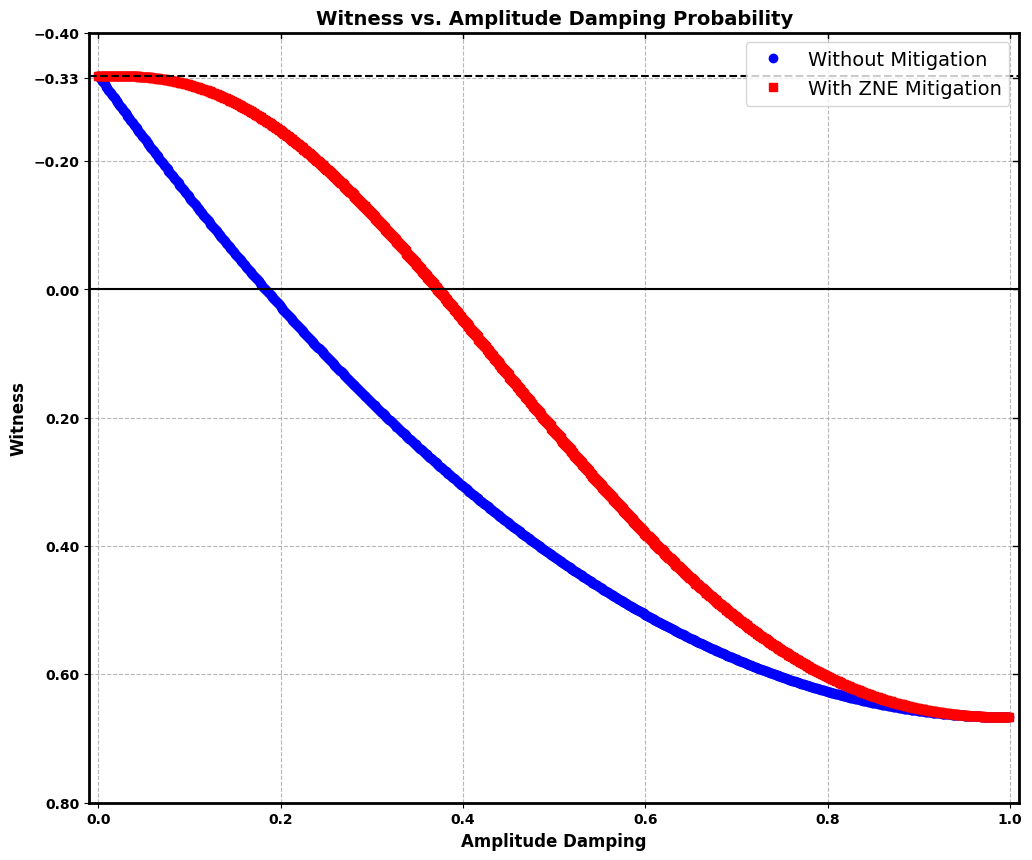}
    \caption{Effect of amplitude damping noise on the Entanglement witness (EW) of Dicke State. The blue curve shows the effect of noise on the EW, whereas the red curve shows the effect of ZNE mitigation method applied to the erroneous state. }
    \label{fig:ad}
\end{figure}

The noise is theoretically modeled using Kraus operators, which describe the evolution of a quantum state under environmental interaction. For a single qubit, the amplitude damping channel is defined by two Kraus operators:
\begin{align}\label{kr_ad}
    E_0 = \begin{pmatrix} 1 & 0 \\ 0 & \sqrt{1-\gamma} \end{pmatrix}, \quad
E_1 = \begin{pmatrix} 0 & \sqrt{\gamma} \\ 0 & 0 \end{pmatrix},
\end{align}

where \(\gamma \in [0,1]\) is the damping probability. The operator \(E_1\) captures the decay of \(\ket{1}\) to \(\ket{0}\) with probability \(\gamma\), while \(E_0\) represents the survival of the \(\ket{1}\) state with amplitude reduced by \(\sqrt{1-\gamma}\). The overall evolution of a single qubit density matrix \(\rho_1\) is given by:
\[
\mathcal{E}(\rho_1) = E_0 \rho_1 E_0^\dagger + E_1 \rho_1 E_1^\dagger.
\]
The parameter \(\gamma\) is linked to the energy relaxation time \(T_1\) via the relation \(\gamma = 1 - e^{-t/T_1}\), where \(t\) is the interaction time \cite{nielsen_chuang_2010}.

To visualize the effect of amplitude damping on the EW of Dicke state, we simulated the corresponding density matrix under noise $\mathcal{E}(\rho_{D^{(2)}_4})$, using \textit{Qiskit's Aer primitive}. We considered local, independent noise applied to each qubit for this purpose, i.e.,  

\begin{align*}
    \noindent \mathcal{E}(\rho_{D^{(2)}_4}) = E_4 \rho_{D^{(2)}_4} E_4^{\dagger},
\end{align*}

where $E_4= \bigotimes_{j \in \{1, 2, 3, 4\}} \sum_{i\in \{1, 2\}} E^j_i $. 

Next, we computed the fidelity between the noisy density matrix $\mathcal{E}(\rho_{D^{(2)}_4})$ and the errorless density matrix $\rho_{D^{(2)}_4}$ of the Dicke state as \cite{nielsen_chuang_2010},

\begin{align*}\label{fdty}
F(\rho_{D^{(2)}_4}, \mathcal{E}(\rho_{D^{(2)}_4})) = \left( \mathrm{Tr}\left[ \sqrt{ \sqrt{\rho_{D^{(2)}_4}} \, \mathcal{E}(\rho_{D^{(2)}_4}) \sqrt{\rho_{D^{(2)}_4}} } \right] \right)^2.    
\end{align*}

Finally, the effect of amplitude damping on the EW of the Dicke state is then computed using Eq.~\eqref{fidelity}. We find that with increasing noise parameter $\gamma$ of the AD noise channel, the four-qubit Dicke state disproportionately degrades, affecting excited-state components, finally leading to loss of entanglement as shown in Fig.~\ref{fig:ad}. Expectedly, this result is in accordance with the existing literature \cite{Campbell_2009}. However, in this work, we further show the impact of error mitigation strategies in detecting the entanglement of the Dicke State.

To analyze the effectiveness of noise mitigation on Markovian amplitude damping noise, we employ \textit{Zero Noise Extrapolation} (ZNE) in Qiskit using Mitiq \cite{mitiq}. It is an error mitigation technique that estimates noise-free expectation values through controlled noise amplification \cite{PhysRevLett.119.180509}. 

For the Markovian amplitude damping channel, ZNE intentionally scales the noise using circuit-level methods like gate repetition. Measurements are taken at multiple noise levels, and with the use of extrapolation models, noiseless expectation values can be inferred \cite{9259940}.

As shown in Fig.~\ref{fig:ad}, ZNE slightly enhances witness values for certain values of \( \gamma \), confirming its utility in partially counteracting damping-induced loss. While effective at moderate noise levels, its efficacy diminishes at high \( \gamma \), where irreversible energy dissipation dominates.

\subsubsection{Non-Markovian Amplitude Damping}

\begin{figure}[h!]
    \centering
    \includegraphics[width=1\linewidth]{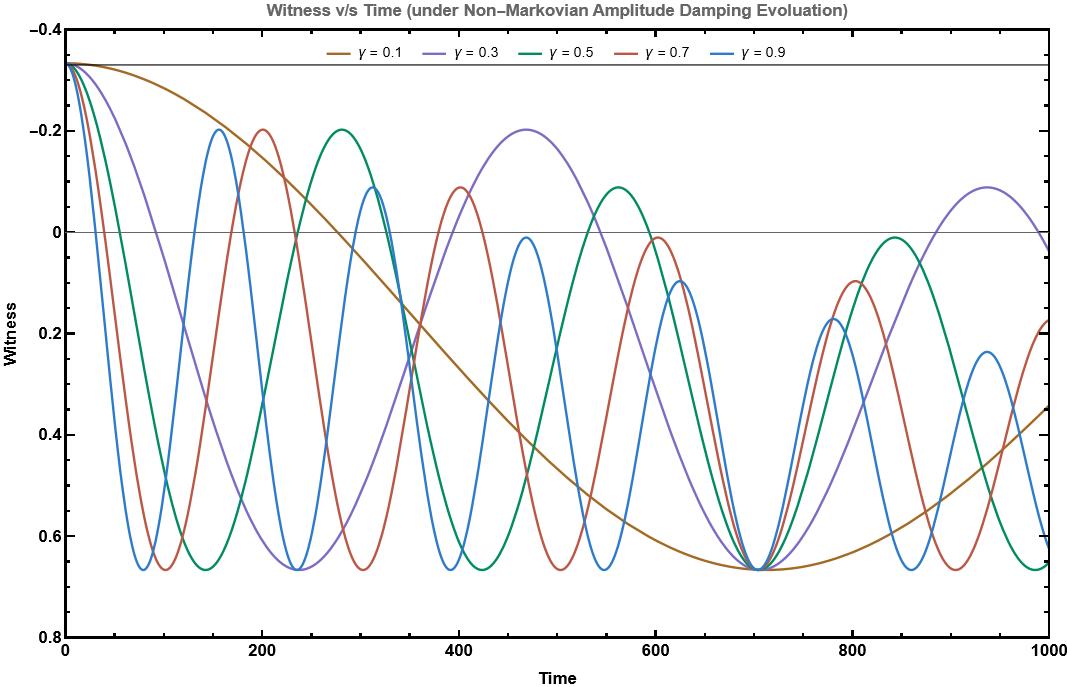}
    \caption{Simulation of effect of non-Markovian amplitude damping channel on EW of Dicke state with $\Gamma = 0.001\,\gamma$.}
    \label{fig:NM_AD}
\end{figure}

Building on the Markovian amplitude damping model, we now consider its non-Markovian counterpart where environmental memory effects become significant. This occurs when reservoir correlation times (\( \tau_r \approx \Gamma^{-1} \)) are comparable to or longer than qubit relaxation times (\( \tau_s \approx \gamma^{-1} \)), violating the Markovian assumption. Here, $\gamma$ represents the coupling strength (identical to the $T_1^{-1}$ rate in the Markovian model), while $\Gamma$ quantifies the reservoir's spectral linewidth. Non-Markovian AD can be described using the following Kraus operators \cite{Paulson_Panwar_Banerjee_Srikanth_2021}:

\[
E_0 = \ket{0}\bra{0} + \sqrt{q}\ket{1}\bra{1}, \quad 
E_1 = \sqrt{1 - q}\ket{0}\bra{1},
\]
With the time-dependent parameter:
\[
q = e^{-\Gamma t} \left[ \cos\left( \frac{d t}{2} \right) + \frac{\Gamma}{d} \sin\left( \frac{d t}{2} \right) \right]^2, \quad 
d = \sqrt{2 \gamma \Gamma - \Gamma^2}.
\]

For our simulation, we assume $\Gamma = 0.001\,\gamma$, which corresponds to a strong memory regime. Under the condition \( \Gamma \ll \gamma \) (i.e., \( d \approx \sqrt{2\gamma\Gamma} \)), the system displays pronounced recoherence oscillations where excitation probability periodically revives due to energy backflow from the environment. This contrasts fundamentally with Markovian amplitude damping's monotonic decay. For the four-qubit Dicke state in our study, these memory effects can lead towards transient entanglement recovery, creating time windows where EW detects stronger quantum correlations than would survive under purely Markovian noise, as shown in Fig.~\ref{fig:NM_AD}.

\subsection{Depolarizing Noise}
\subsubsection{Markovian Depolarizing Noise}
Markovian depolarizing noise is a common type of quantum noise that introduces random errors by mixing a quantum state with a maximally disordered state. It symmetrically affects all components of a qubit’s state, causing both amplitude and phase errors. Unlike amplitude damping (which targets energy loss), depolarizing noise represents a uniform degradation of quantum information, pushing states toward complete randomness. This makes it a critical error model for studying the robustness of quantum algorithms and entanglement in noisy environments.

The Markovian depolarizing channel is described using Kraus operators :
\begin{align}
\nonumber \noindent & K_0 = \sqrt{1-p} \, I, \quad
K_1 = \sqrt{\frac{p}{3}} \, X,\\ \quad
& K_2 = \sqrt{\frac{p}{3}} \, Y, \quad \quad
K_3 = \sqrt{\frac{p}{3}} \, Z,   
\end{align}

where \( p \in [0,1] \) is the depolarizing probability, \( I \) is the identity operator, and \( X, Y, Z \) are Pauli operators. The channel transforms a density matrix \( \rho \) into:
\[
\mathcal{E}(\rho) = K_0 \rho K_0^\dagger + K_1 \rho K_1^\dagger + K_2 \rho K_2^\dagger + K_3 \rho K_3^\dagger.
\]
Here, \( K_0 \) preserves the original state with probability \( 1-p \), while \( K_1, K_2, K_3 \) introduce random bit-flip (\( X \)), phase-flip (\( Z \)), or combined (\( Y \)) errors, each occurring with probability \( p/3 \). For \( n \)-qubit systems, noise may act independently on each qubit (local depolarizing) or globally replace the entire state with the maximally mixed state \( I^{\otimes n}/2^n \) with probability \( p \) \cite{nielsen_chuang_2010}.

\begin{figure}[hbt!]
    \centering
    \includegraphics[width=1\linewidth]{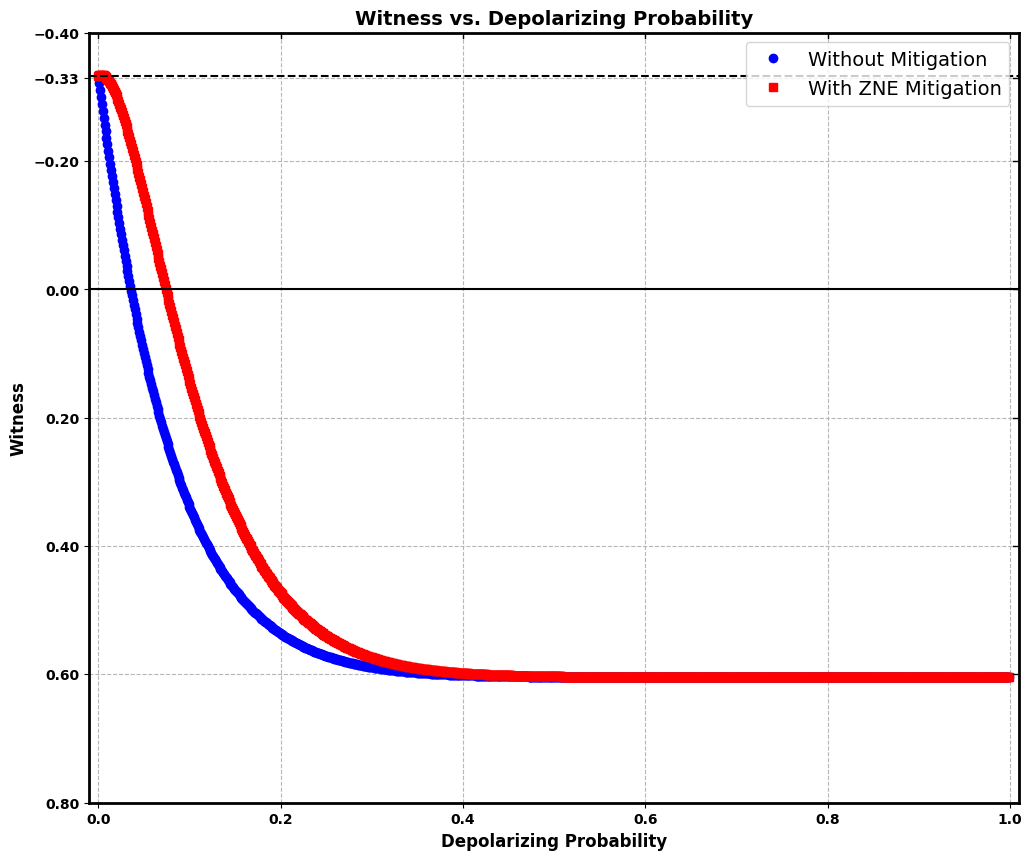}
    \caption{Simulation of the effect of depolarizing probability on the EW of the Dicke state with the effect of the ZNE mitigation method.}
    \label{fig:depol}
\end{figure}

Depolarizing noise reduces the coherence of any state in a superposition by randomizing qubit states. A superposition \( \alpha\ket{0} + \beta\ket{1} \) loses purity, evolving into a mixed state with diminished quantum properties. For example, GHZ states exhibit exponential fidelity decay \( F \approx (1 + (1-p)^n)/2 \) with qubit count \( n \), while W states show relative resilience due to distributed entanglement \cite{ali-2014}. Depolarizing noise affects the four-qubit Dicke state by degrading the state and correlations as a result of noise disruptions due to random Pauli errors. As a result, even small \( p \) significantly reduces fidelity, as shown in Fig.~\ref{fig:depol}.

When applying ZNE to Markovian depolarizing noise, for lower values of \( p \), a slight improvement is seen. However, as the probability of depolarizing increases, ZNE makes no difference as seen in Fig.~\ref{fig:depol}. The divergence from amplitude damping (where ZNE worked to higher \( \gamma \)) occurs because depolarizing noise symmetrically destroys all quantum information, while amplitude damping preserves ground-state components. Method of simulation of the effect of Markovian depolarizing noise on Dicke state, calculation of EW, and simulation of the effect of ZNE remains the same as in the AD case.

\subsubsection{Non-Markovian Depolarizing Noise}

\begin{figure}[h!]
    \centering
    \includegraphics[width=1\linewidth]{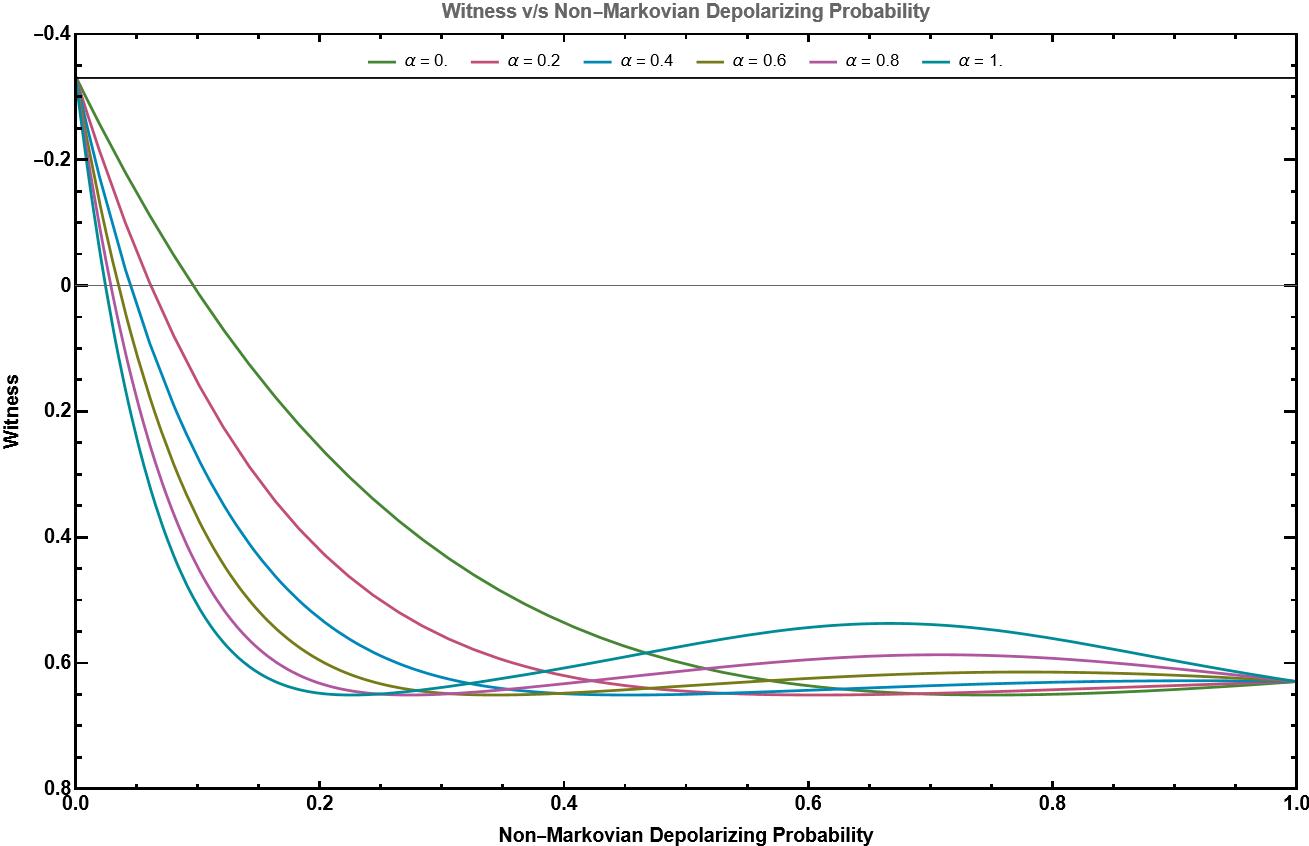}
    \caption{Simulation of the effect of a non-Markovian depolarizing channel on the EW of the Dicke state for different values of $\alpha$.}
    \label{fig:nm_depol}
\end{figure}
Building on the Markovian depolarizing model, which symmetrically degrades quantum states through random Pauli errors with fixed probability \( p \), the non-Markovian variant incorporates environmental memory effects via the parameter \( \alpha \). This memory parameter \( \alpha \geq 0 \) quantifies the degree of correlations with the environment: when \( \alpha > 0 \), past interactions influence current error rates, enabling information backflow from the environment to the qubit~\cite{PhysRevA.98.032328,PhysRevA.110.052209}.

Mathematically, this is encoded in modified Kraus operators:
\[
\begin{aligned}
K_I(p) &= \sqrt{(1 - 3\alpha p)(1 - p)} \, I, \\
K_X(p) &= \sqrt{\left[1 + 3\alpha(1 - p)\right] \frac{p}{3}} \, X, \\
K_Y(p) &= \sqrt{\left[1 + 3\alpha(1 - p)\right] \frac{p}{3}} \, Y, \\
K_Z(p) &= \sqrt{\left[1 + 3\alpha(1 - p)\right] \frac{p}{3}} \, Z.
\end{aligned}
\]

Here, \( \alpha \) explicitly couples the error dynamics to historical evolution: the factor \( (1 - 3\alpha p) \) in \( K_I \) enhances state preservation during memory-retention periods, while \( [1 + 3\alpha(1 - p)] \) in \( K_{X/Y/Z} \) modulates error injection. This contrasts with Markovian noise (\( \alpha = 0 \)), where error accumulation is memoryless and irreversible. Crucially, \( \alpha \) represents the non-Markovianity strength: higher values correlate with longer environmental memory timescales, allowing transient suppression of depolarization and revival of entanglement, as shown in Fig.~\ref{fig:nm_depol}. To ensure the complete positivity of the complete dynamic map, we assumed both $p$ and $\alpha \in [0, 1]$ \cite{PhysRevA.110.052209}.

\subsection{Bit-Phase Flip Error}

A bit-phase flip error is a quantum error that combines a bit flip (swapping \( \ket{0} \) and \( \ket{1} \)) and a phase flip (inverting the sign of the \( \ket{1} \) component). Represented by the Pauli-\( Y \) operator, it acts on a qubit state \( \alpha\ket{0} + \beta\ket{1} \) as:
\[
Y = \begin{pmatrix} 0 & -i \\ i & 0 \end{pmatrix}, \quad Y(\alpha\ket{0} + \beta\ket{1}) = -i\beta\ket{0} + i\alpha\ket{1}.
\]
This operation flips the qubit’s basis states (bit flip) and introduces a relative phase shift (phase flip), disrupting both amplitude and phase coherence. On the Bloch sphere, it corresponds to a 180° rotation around the Y-axis \cite{nielsen_chuang_2010}.

\begin{figure}[h!]
    \centering
    \includegraphics[width=1\linewidth]{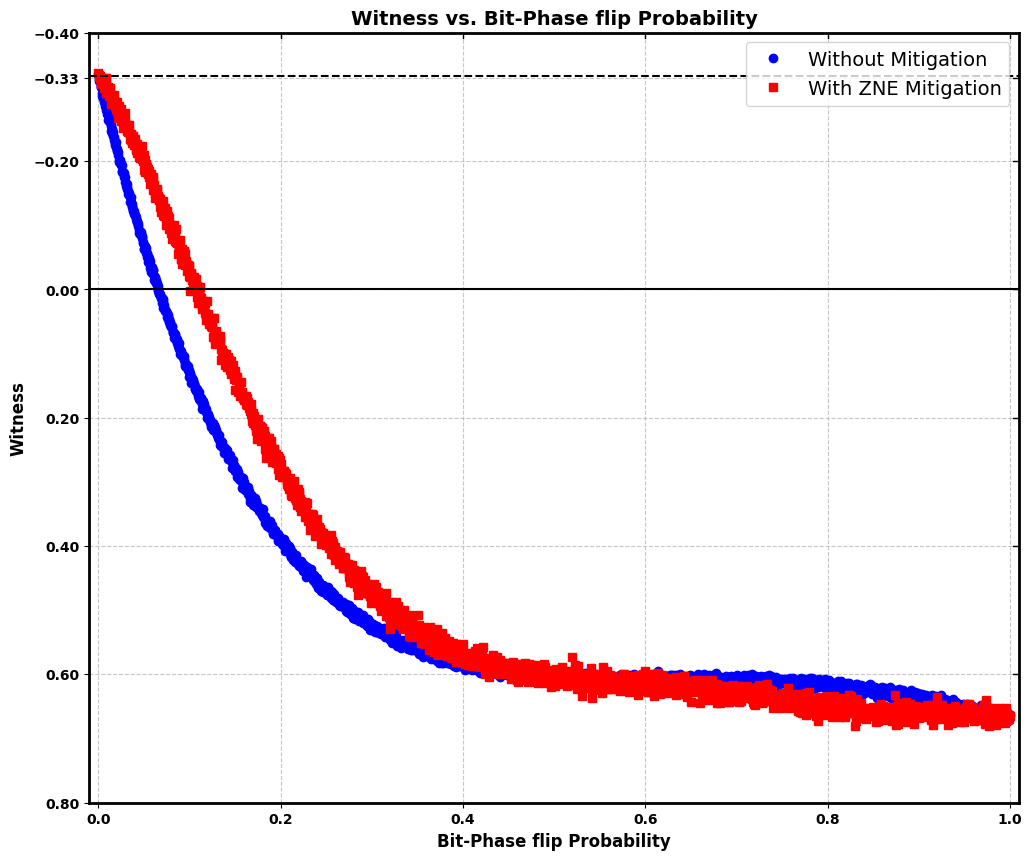}
    \caption{Simulation of the effect of Bit-Phase flip error probability on the EW of the Dicke state with the effect of the ZNE mitigation method.}
    \label{fig:bit-phase}
\end{figure}

Bit-phase flips degrade superposition states and entangled systems. For example, a Bell state \( (\ket{00} + \ket{11})/\sqrt{2} \) subjected to a \( Y \)-error on one qubit becomes \( (\ket{01} - \ket{10})/\sqrt{2} \), breaking entanglement. Crosstalk between qubits can introduce correlated \( Y \)-errors, reducing algorithm fidelity.

Bit-phase flips are a subset of depolarizing noise, which includes all Pauli errors (X, Y, Z). Unlike amplitude damping (\( T_1 \) noise), which asymmetrically drives \( \ket{1} \rightarrow \ket{0} \), \( Y \)-errors symmetrically disrupt both populations and phases. They are more disruptive than pure dephasing (\( T_2 \) noise), as they scramble both computational and phase information.

When applying ZNE to Bit-Phase flip error, for lower values of \( p \), a slight improvement is seen. However, as the probability of error increases, ZNE makes no difference as seen in Fig.~\ref{fig:bit-phase}.

\subsection{Readout Error}

Readout errors (or measurement errors) occur when the process of measuring a qubit’s state misidentifies its true value. Unlike noise during computation (e.g., depolarizing or amplitude damping), these errors arise at the final measurement stage, corrupting classical outcomes. For instance, a qubit in state \( \ket{0} \) might be read as \( \ket{1} \) with probability \( \epsilon \), and vice versa. Readout errors represent classical noise in quantum-to-classical conversion, distinct from quantum decoherence. While not altering the quantum state itself, they distort the observed outcomes, limiting algorithm accuracy.

\begin{figure}[hbt!]
    \centering
    \includegraphics[width=1\linewidth]{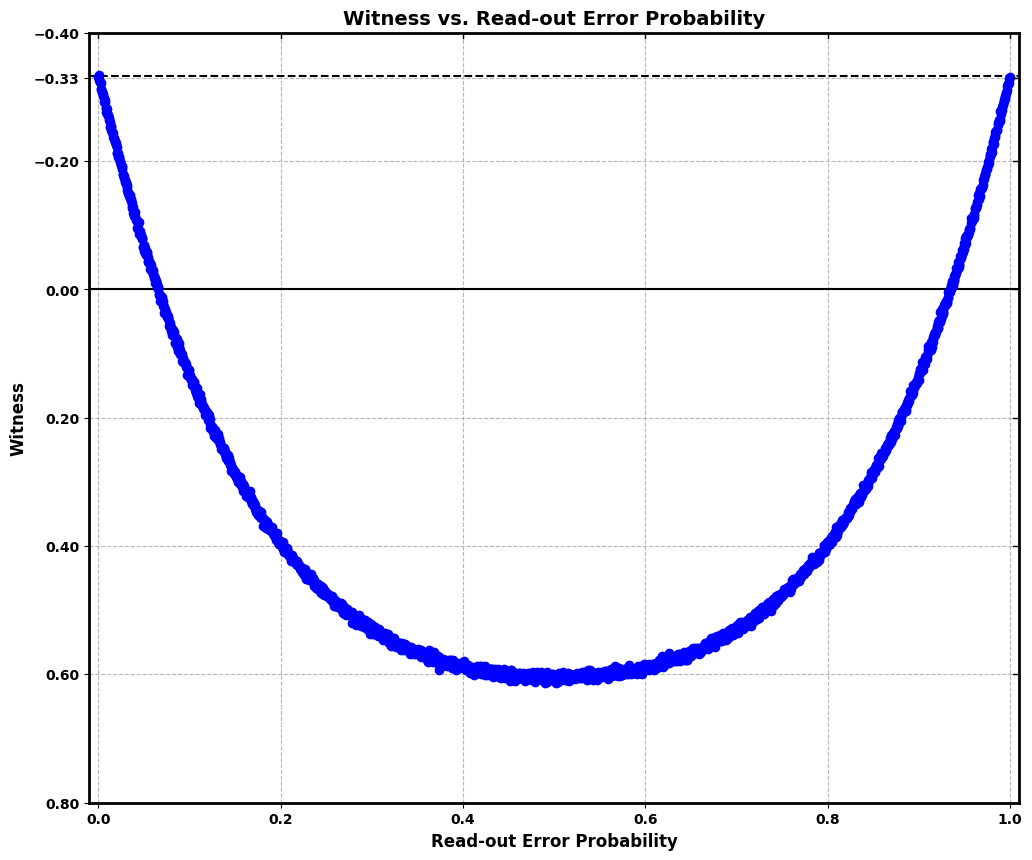}
    \caption{Simulation of the effect of read-out error probability on EW of the Dicke state.}
    \label{fig:readout}
\end{figure}

Readout errors are modeled using a \textbf{confusion matrix} \( A \). For \( n \)-qubit systems, \( A \) is a \( 2^n \times 2^n \) matrix where each entry \( A(y|x) \) represents the probability of measuring outcome \( y \) when the true state is \( x \). For a single qubit, this simplifies to:
\[
A = \begin{pmatrix} 
1-\epsilon_0 & \epsilon_1 \\ 
\epsilon_0 & 1-\epsilon_1 
\end{pmatrix},
\]
where \( \epsilon_0 \) (\( \epsilon_1 \)) is the error rate for misreading \( \ket{0} \) as \( \ket{1} \) (\( \ket{1} \) as \( \ket{0} \)) \cite{PhysRevApplied.23.024055}. The measured probabilities \( p_{\text{meas}} \) relate to the true probabilities \( p_{\text{true}} \) via:
\[
p_{\text{meas}} = A \cdot p_{\text{true}}.
\]

Readout errors can degrade observed entanglement. For example, a Bell state \( (\ket{00} + \ket{11})/\sqrt{2} \) may be misread as \( \ket{01} \) or\( \ket{10} \), reducing fidelity and masking genuine quantum correlations.
For the Dicke state, as the error probability \( p \) increases, the state loses quantum correlations; however, after the error probability \( p \) increases beyond 0.5, revival of quantum correlation can be seen, as depicted in Fig.~\ref{fig:readout}. We conjecture that this revival is due to the inherent inversion (bit-reversal) symmetry of the Dicke state. With increasing noise parameters, the probability of \textit{all qubits} to be read wrong (flipped) increases significantly, taking the Dicke state back to its original state, and thus providing an inherent robustness against read-out error. The symmetric nature of this revival is due to our assumption of all four noise channels having same noise parameter. Methods and tools of simulation to evaluate the effect of readout error on the EW of the Dicke state remain the same as those done for the Markovian AD.

\subsection{Thermal Relaxation}

For the state-of-the-art superconducting quantum processors, \( T_1 \) and \( T_2 \) times are critical benchmarks defining computational reliability. \( T_1 \) (energy relaxation time) measures how long a qubit retains its energy before decaying from the excited state (\( \ket{1} \)) to the ground state (\( \ket{0} \)), a process linked to AD. \( T_2 \) (coherence time) reflects the duration a qubit maintains phase coherence in superposition states (\( \alpha\ket{0} + \beta\ket{1} \)) before environmental noise disrupts quantum interference, connecting to dephasing and depolarizing effects. 

\begin{figure}[h!]
    \centering
    \includegraphics[width=1\linewidth]{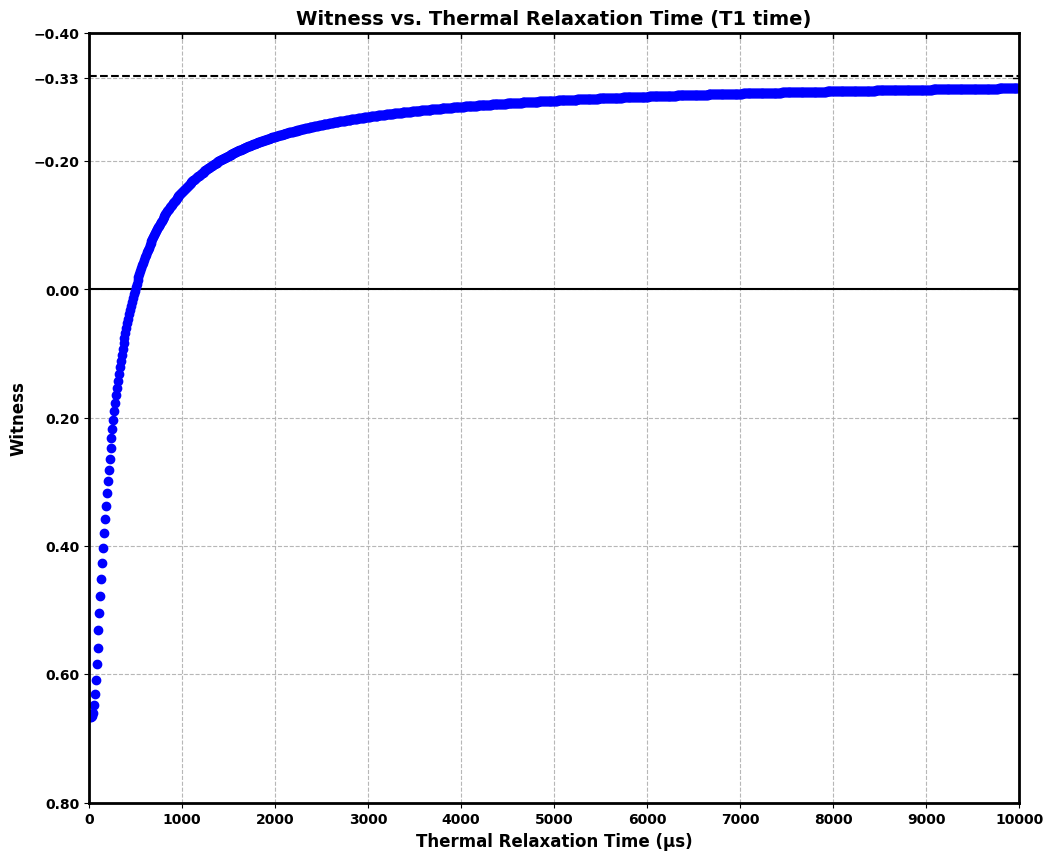}
    \caption{Simulation of the Effect of thermal relaxation time on EW.}
    \label{fig:thermal}
\end{figure}

These times directly limit the number of error-tolerant operations a quantum processing unit (QPU) can perform \cite{kandala-2017}. Longer \( T_1 \) and \( T_2 \) reduce gate error rates and enhance the fidelity of entangled states, as shown in Fig.~\ref{fig:thermal}. The simulation methods and tools used to assess the impact of readout error on the EW of the Dicke state are the same as those applied in the Markovian AD case. As can be seen from Fig.~\ref{fig:thermal}, the EW of the Dicke state saturated around 6000 microseconds of \( T_1 \) time. This roughly translates to the following: on a QPU, to perform any meaningful tasks using the Dicke state, at least 6000 microseconds \( T_1 \) time per qubit is required. 

\section{Conclusion}\label{conc}
In this work, we have theoretically constructed an Entanglement Witness (EW) for the four-qubit Dicke state, which has the maximum negative values of $-0.33$ for the Dicke state and positive values for every element in the set of all separable states. We show experimental generation of the four-qubit Dicke state using two methods and verify EW on the two different IBM QPUs named \texttt{ibm\_sherbrook} and \texttt{ibm\_brisbane}. For the statevector-based method of Dicke state preparation, we find that the EW attains a negative value even without any error mitigation method, implying that the state remains globally entangled. Further, we explore various error mitigation methods, such as DD, TREX, and T, and show that with error mitigation, the EW of the Dicke State improves significantly. In contrast, with the unitary-based method of state preparation, the EW attains negative values solely if one uses error mitigation methods. This is because this method uses a higher number of quantum gates at the native level implementation (transpiled circuit) than the statevector-based method, which introduces more noise in the system. This result is of particular interest in the present era of quantum utility, as we show even with today's $127-$ qubit quantum computers, achieving the theoretical global entanglement of a $4-$ qubit quantum state is significantly difficult, especially without error mitigation. 

We have also theoretically analyzed the impact of various noise channels on the entanglement witness (EW) of the four-qubit Dicke state, including Markovian and non-Markovian amplitude damping (AD) and depolarizing noise, bit-phase flip errors, readout errors, and thermal relaxation. Our results indicate that the EW is highly sensitive to the parameters of these noise models. In the case of Markovian AD and Markovian depolarizing noise, the EW vanishes entirely beyond noise probability, p = 0.185 and p = 0.037, respectively. We further explored the application of zero-noise extrapolation (ZNE) as a mitigation strategy, which yielded a modest improvement in the EW compared to the unmitigated case, as EW vanishes beyond noise probability, p = 0.374 and p = 0.076, respectively. For non-Markovian AD and depolarizing channels, we observed a partial revival of the EW tied to the parameter of non-Markovianity, highlighting the influence of memory effects in preserving entanglement. Interestingly, under readout errors, we detected a revival of the EW with increasing error probability. As readout error probability increases beyond p = 0.067, EW vanishes; however, beyond p = 0.935, entanglement revival is seen. This behavior appears to be linked to the inherent bit-reversal symmetry of the Dicke state.
Additionally, we investigated the impact of bit-phase flip noise, where the EW became positive at relatively low error probabilities (p = 0.0675), indicating a loss of detectable entanglement. However, the application of mitigation techniques enabled the slight recovery of the EW signal, with EW becoming positive at p = 0.111, indicating loss of entanglement. In the case of thermal relaxation, the EW gradually emerges as the qubits relaxed (T1 = 490 $\mu$s), eventually reaching a steady-state saturation near its maximal value over longer timescales.
\vspace{5mm}

\section*{Data Availability statement}
All data that support the findings of this study are included in the article.

\section*{Acknowledgement}
VN acknowledges the support from the Interdisciplinary Cyber Physical Systems (ICPS) programme of the Department of Science and Technology (DST), India, Grant No.: DST/ICPS/QuST/Theme-1/2019/6.
Tomis Prajapati would like to acknowledge CSIR for the research funding. 

\begin{widetext}

\appendix

\section*{Appendix}

\section{Calibration Data of utilized QPUs}

Calibration data of \textit{ibm\_sherbrooke} obtained from the IBM quantum platform during one of the executions on 14 June 2025 at 07:44:41 UTC. For the four-qubit Dicke state, the utilized qubits were q\_40, q\_41, q\_42, and q\_53.

\begin{table*}[h]
\centering
\begin{tabular}{|c|c|c|c|c|c|l|}
\hline
\textbf{Qubit} & \textbf{T1 ($\mu$s)} & \textbf{T2 ($\mu$s)} & \textbf{Frequency (GHz)} & \textbf{Readout error} & \textbf{Pauli-X error} & \textbf{ECR error} \\
\hline
q\_40 & 386.539340 & 522.0580825 & 4.70550458024523 & 0.00634765625 & 0.0000900317404829324 & ECR(40\_41): 0.010306460451835947 \\
\hline
q\_41 & 133.616765 & 196.2051255 & 4.814544587966411 & 0.01123046875 & 0.0004680582600952883 &
\makecell[l]{ECR(41\_40): 0.010306460451835947 \\
ECR(41\_42): 0.0146881946910693 \\
ECR(41\_53): 0.007693905874916834} \\
\hline
q\_42 & 412.0483456 & 571.113260 & 4.654685942674868 & 0.047119140625 & 0.0008706416620557398 & ECR(42\_41): 0.0146881946910693 \\
\hline
q\_53 & 300.210155 & 375.184213 & 4.7677893765194925 & 0.0869140625 & 0.00018240250522785623 & ECR(53\_41): 0.007693905874916834 \\
\hline
\end{tabular}
\caption{Calibration data from \textit{ibm\_sherbrooke} for Physical Qubits used for execution of witness test for  Dicke State. Including relaxation and dephasing times, frequencies, readout errors, Pauli-X gate errors, and ECR (two-qubit conditional gate) gate errors.}
\label{tab:calibration_sherbrooke}
\end{table*}

Calibration data of \textit{ibm\_brisbane} obtained from the IBM quantum platform during one of the executions on 11 June 2025 at 05:21:09 UTC. For the four-qubit Dicke state, the utilized qubits were q\_57, q\_58, q\_59, and q\_71.

\begin{table*}[h]
\centering
\begin{tabular}{|c|c|c|c|c|c|l|}
\hline
\textbf{Qubit} & \textbf{T1 ($\mu$s)} & \textbf{T2 ($\mu$s)} & \textbf{Frequency (GHz)} & \textbf{Readout error} & \textbf{Pauli-X error} & \textbf{ECR error} \\
\hline
q\_57 & 260.010343 & 114.4587787 & 4.927642840200743 & 0.028564453125 & 0.000809123458704459 & ECR(57\_58): 0.005037455569564864 \\
\hline
q\_58 & 199.7212230 & 142.6440762 & 4.8869796930882705 & 0.012939453125 & 0.0001389173346 & 
\makecell[l]{ECR(58\_57): 0.005037455569564864 \\ 
ECR(58\_59): 0.004594822610177379 \\
ECR(58\_71): 0.004989295274236055} \\
\hline
q\_59 & 175.9719413 & 61.9571231 & 4.971230281148097 & 0.018798828125 & 0.00023781172279594302 & 
ECR(59\_58): 0.004594822610177379 \\
\hline
q\_71 & 317.3842217 & 345.2441725 & 4.788274139812204 & 0.00830078125 & 0.00013051150052018743 & ECR(71\_58): 0.004989295274236055 \\
\hline
\end{tabular}
\caption{Calibration data from \textit{ibm\_brisbane} for Physical Qubits used for execution of witness test for  Dicke State. Including relaxation and dephasing times, frequencies, readout errors, Pauli-X gate errors, and ECR (two-qubit conditional gate) gate errors.}
\label{tab:calibration_brisbane}
\end{table*}
\end{widetext}

\bibliography{ref}
\end{document}